\documentclass[%
reprint,
superscriptaddress,
 amsmath,amssymb,
 aps,
pra,
]{revtex4-2}

\usepackage{graphicx}
\usepackage{dcolumn}
\usepackage{bm}
\usepackage{hyperref}
\usepackage{tabularx,booktabs}
\usepackage{soul}
\usepackage{amsmath}
\usepackage{braket}
\hypersetup{
    colorlinks=true,
    linkcolor=black,
    citecolor=blue,
    urlcolor=black
    }

\usepackage{color}
\usepackage{orcidlink}

\usepackage{algorithm} 
\usepackage{algpseudocode} 

\usepackage{amsmath}

\begin{document}

\preprint{APS/123-QED}

\title{Reconfigurable Four-Photon Interference among Three Nodes on a  Field Deployed Metropolitan Fiber Network \\
}

\author{Kazi Reaz\orcidlink{0000-0001-9106-4736}}
\affiliation{Department of Physics \& Astronomy, University of Tennessee, Chattanooga, TN 37403, USA}
\affiliation{UTC Quantum Center, University of Tennessee, Chattanooga, TN 37403, USA}

\author{Md Mehdi Hassan\orcidlink{0000-0002-3643-8229}}
\affiliation{Department of Physics \& Astronomy, University of Tennessee,  Knoxville, TN 37996, USA}

\author{Jacob E. Humberd}
\affiliation{Department of Physics \& Astronomy, University of Tennessee,  Chattanooga, TN 37403, USA}

\author{Matthew L. Boone}
\affiliation{Department of Physics \& Astronomy, University of Tennessee,  Chattanooga, TN 37403, USA}
\affiliation{Department of Physics \& Astronomy, University of Tennessee, Knoxville, TN 37996, USA}

\author{Angel Fraire Estrada}
\affiliation{Department of Physics \& Astronomy, University of Tennessee, Chattanooga, TN 37403, USA}

\author{Rick Mukherjee}
\affiliation{Department of Physics \& Astronomy, University of Tennessee, Chattanooga, TN 37403, USA}
\affiliation{UTC Quantum Center, University of Tennessee, Chattanooga, TN 37403, USA}

\author{H. R. Sadeghpour\orcidlink{0000-0001-5707-8675}}
\affiliation{ITAMP, Center for Astrophysics $|$ Harvard \& Smithsonian, Cambridge, MA 02138, USA}

\author{Girish S. Agarwal\orcidlink{0000-0002-8320-4708}}
\affiliation{Institute for Quantum Science and Engineering, Department of Physics \& Astronomy, Department of Biological and Agricultural Engineering,
Texas A\&M University, College Station, TX 77843, USA}

\author{George Siopsis\orcidlink{0000-0002-1466-2772}}
\affiliation{Department of Physics \& Astronomy, University of Tennessee, Knoxville, TN 37996, USA}

\author{Tian Li\orcidlink{0000-0003-2993-0386}}
\email{tian-li@utc.edu}
\affiliation{Department of Physics \& Astronomy, University of Tennessee, Chattanooga, TN 37403, USA}
\affiliation{UTC Quantum Center, University of Tennessee, Chattanooga, TN 37403, USA}


\begin{abstract}

\textcolor{black}{Advanced quantum networking protocols beyond bi-photon, point-to-point links rely critically on the ability to perform multi-photon interference across multiple nodes under realistic operating conditions. Yet experimental validation of such higher-order, multi-node interference effects in deployed metropolitan fiber networks remains limited. Here, we report a field demonstration of polarization-controlled reconfigurable four-photon interference over three distant nodes on a deployed metropolitan fiber network. Using a fully fiber-coupled linear-optical platform, we observe a fusion-type four-photon interference signature in presence of real-world impairments, including photon loss, polarization drift, and timing uncertainty. By performing polarization-resolved measurements on two locally retained photons, we conditionally select distinct two-photon coincidence channels that exhibit Bell-like and N00N-like behavior. Rather than pursuing multi-partite entanglement verification, this work focuses on establishing the technical feasibility of multi-photon, multi-node interference and reconfigurable conditional state preparation in the field in a deployed fiber network environment. These results serve as a systems-level validation toward future multi-photon, multi-node quantum networking architectures that require robust interference performance outside the laboratory.}

\end{abstract}

\maketitle


\section{Introduction}

Quantum networks, with the ultimate vision of a quantum Internet, are poised to enable the transmission of quantum states and nonclassical correlations between spatially separated resources. Such capabilities are foundational for secure communication, distributed quantum computation, and networked quantum sensing beyond what is achievable with purely classical signals~\cite{Pirandola2016,wehner2018quantum}. Over the past 5 years, most quantum networking research on \textit{metro-scale fiber networks} worldwide has emphasized the robust distribution and verification of \textit{bi-photon polarization entanglement} over deployed telecommunication infrastructure, enabled by active polarization compensation, high-efficiency superconducting detection, and precise timing synchronization. For example, polarization entanglement from a type-II spontaneous parametric down-conversion (SPDC) source in the telecom O-band was distributed over 10~km of deployed fiber with active polarization stabilization in Singapore~\cite{shi2020stable}. Researchers in Austria and Slovakia transmitted polarization-entangled SPDC photons across 248~km of deployed fiber, achieving sustained high visibility using automated polarization control and nonlocal dispersion compensation~\cite{neumann2022continuous}. In New York City, polarization-entangled photon pairs at 1324~nm–-795~nm generated via four-wave mixing in warm rubidium vapor were distributed across a 34~km network with high fidelity and uptime~\cite{craddock2024automated}. In China, polarization-entangled photon pairs generated by a silicon photonic chip via spontaneous four-wave mixing were distributed over 30~km, with a CHSH Bell violation exceeding 27.8~$\sigma$~\cite{jiang2025entanglement}. Additional metro-scale demonstrations have employed quantum dots for polarization-correlation distribution and related timing applications~\cite{alqedra2025entanglement,shooter20201ghz}. Most significantly, researchers in Chicago employed polarization entanglement generated via type-0 cascaded second-harmonic generation SPDC to demonstrate quantum teleportation over 30.2~km of fiber co-propagating high-power classical telecom traffic~\cite{thomas2024quantum}, validating quantum--classical coexistence in shared infrastructure through optimized transmission bands and noise-mitigation techniques.

\begin{figure*}
\centering
\includegraphics[width=\linewidth]{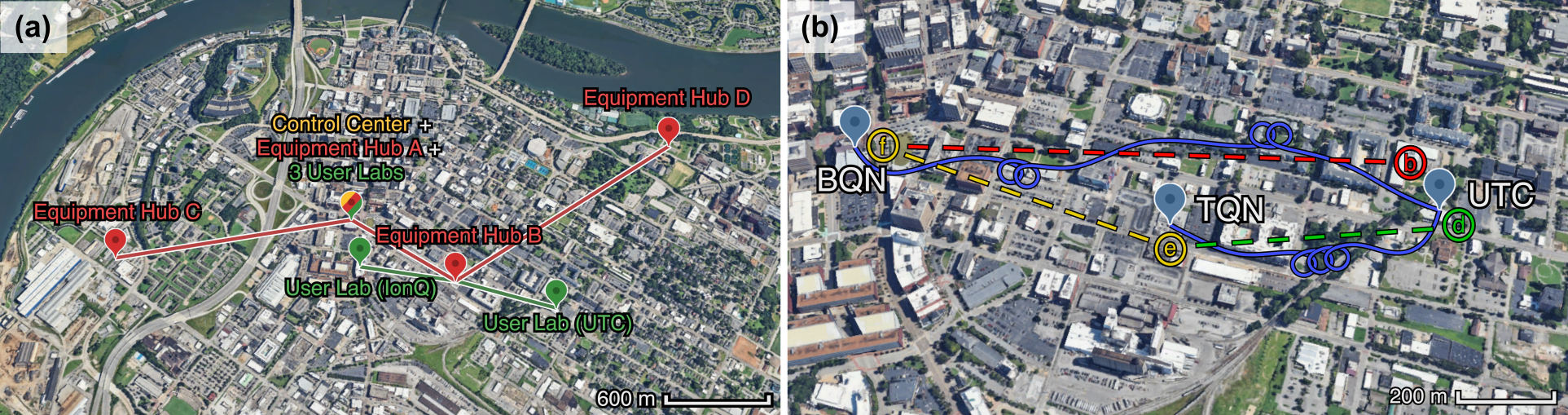}
\caption{A bird’s-eye view of the EPB quantum network architecture in downtown Chattanooga is shown in (a)~~\cite{earl2022architecture}, with the segment utilized in this work highlighted in (b). Since Equipment Hubs A and B are located on Broad Street and Tenth Street, respectively, we label them as BQN and TQN. The deployed fiber loops between UTC-TQN and UTC-BQN are approximately 5~km and 10~km in length respectively. The interference of the four photons labeled $b$, $d$, $e$, and $f$, generated by the scheme shown in Fig.~\ref{fig:Layout}, is distributed among the three distant nodes ($b$ and $d$ are retained at UTC, while $e$ and $f$ are distributed to TQN and BQN, respectively) of the deployed fiber network. 
\label{fig:Birdview}}
\end{figure*}

\textcolor{black}{While these demonstrations establish that deployed fiber networks can support high-fidelity \textit{bi-photon} entanglement distribution across \textit{2-node, point-to-point} links at metropolitan scales, extending such experiments to \textit{multi-photon interference among muliple nodes}, which underpins many advanced quantum-network protocols~\cite{park2021four,huang2011experimental,pan2001experimental}, over deployed networks remains a significant challenge. Multi-photon experiments over multiple nodes are substantially more sensitive to loss, background noise, polarization drift, and timing uncertainty, and are therefore often restricted to laboratory settings using fiber spools or short links~\cite{rubeling2025fiber,lo2023generation,pompili2021realization,fang2019three}. Nevertheless, developing and validating field-ready methods for \textit{multi-photon interference across multiple nodes} is an important intermediate step toward scalable quantum-network protocols that rely on higher-order correlations, multiplexing, and conditional heralding.}

In this work, we demonstrate polarization-controlled reconfigurable \textit{four-photon interference} among \textit{three distant nodes} over a commercially deployed, metropolitan-scale telecom C-band fiber network. Using an all-fiber platform composed of a fully fiber-coupled bi-photon source and linear-optical fusion based on fiber beam splitters, we observe four-photon coincidence interference under temporal overlap. We use polarization-resolved local measurements as a reconfigurable control knob to select between distinct \textit{conditional coincidence channels with Bell-like and N00N-like behaviors} associated with different multi-photon interference pathways. This functionality is presented as \textit{a network control and benchmarking primitive}, enabling systematic exploration of photon loss sensitivity (with N00N states amplifying photon loss and distinguishability errors), polarization drift robustness, and interference stability in a deployed environment, rather than as a final application or a demonstration of entanglement distribution.


\textcolor{black}{The significance and novelty of this work are twofold: (1) it demonstrates a reconfigurable, polarization-controlled four-photon interference protocol among three distant nodes over real-world telecommunication infrastructure, operating under substantial loss and environmental perturbations; and (2) it extends photon interference geometries from two-photon, point-to-point configurations—implemented either locally~\cite{wang2019research,ma2015hong} or across a network~\cite{sun2016quantum,sun2017entanglement,davis2025entanglement}—to a four-photon, three-node configuration realized simultaneously. \textit{We explicitly focus on interference- and correlation-based signatures rather than entanglement verification}, and we discuss the technical improvements required to enable rigorous multi-partite entanglement witnesses in future experiments.}


The paper is arranged as follows. We begin with an overview of the deployed
metropolitan fiber network infrastructure in downtown Chattanooga, followed by a theoretical model of the four-photon fusion stage and derive the polarization-dependent conditional coincidence structure corresponding to Bell-like and N00N-like correlation channels. We then describe the experimental methodology, including characterization of the fully fiber-coupled bi-photon source, synchronization of photon arrival times across the metro-scale network, and implementation of the fusion protocol. Next, we present field measurements demonstrating polarization-controlled reconfigurable four-photon interference and conditional remote two-photon coincidence signatures across spatially separated network nodes. Finally, we analyze the dominant performance limitations in the current field configuration and outline a pathway toward rigorous entanglement verification enabled by forthcoming upgrades to network hardware and source fidelity.


\section{Overview of the deployed fiber network}

EPB, a municipal electric utility and fiber-optic communication provider, owns and operates a deployed quantum-network testbed in downtown Chattanooga~\cite{earl2022architecture}. Designed to support photonic qubit generation, distribution, and measurement across multiple nodes, the network enables researchers from industry, government, and academia to test devices and protocols in a deployed fiber-optic environment. 
A bird’s-eye view of the entire network is shown in Fig.~\ref{fig:Birdview}(a), with the segment utilized in this work highlighted in Fig.~\ref{fig:Birdview}(b). In this work, the four-photon interference is shared among the three distant nodes shown in Fig.~\ref{fig:Birdview}(b), with two photons retained at one node and the other two distributed to the remaining nodes over the deployed fiber network.


\section{Theoretical Models}

\subsection{Four-Photon Fusion Stage and Conditional Projection Channels}


\textcolor{black}{In the degenerate type-II spontaneous parametric down-conversion (SPDC) process used in this work, the nonlinear crystal emits photon pairs with orthogonal polarizations into the same spatial mode. To lowest order in the pair-generation probability, the emitted two-photon state can be written in operator form as
\begin{equation}
|\psi\rangle_{\text{pair}} \propto a_H^\dagger a_V^\dagger |0\rangle ,
\end{equation}
where $a_H^\dagger$ and $a_V^\dagger$ are creation operators for horizontally and vertically polarized photons in spatial mode $a$. In this form the photons are not yet spatially separated, and therefore the state does not immediately correspond to a polarization-entangled Bell pair.}

\textcolor{black}{In our experiment, spatial separation of the photons is achieved probabilistically using a sequence of 50:50 fiber beam splitters. When the initial spatial mode $a$ is incident on a beam splitter (BS), the transformation of the creation operators is given by
\begin{equation}
a_p^\dagger \rightarrow \frac{1}{\sqrt{2}} \left( t_p^\dagger + r_p^\dagger \right),
\label{eqn:spatial}
\end{equation}
where $p \in \{H,V\}$ denotes polarization, and $t$ and $r$ represent the two output spatial modes. Equation~(\ref{eqn:spatial}) indicates that both polarizations reside within the same spatial mode. 
}

\textcolor{black}{Applying this transformation to the two-photon state yields
\begin{align}
a_H^\dagger a_V^\dagger |0\rangle
&\rightarrow
\frac{1}{2}
(t_H^\dagger + r_H^\dagger)(t_V^\dagger + r_V^\dagger)|0\rangle \\
&=
\frac{1}{2}\Big(
t_H^\dagger t_V^\dagger
+
t_H^\dagger r_V^\dagger
+
r_H^\dagger t_V^\dagger
+
r_H^\dagger r_V^\dagger
\Big)|0\rangle.
\end{align}
The first and last terms correspond to two photons exiting the same port of the beam splitter. In the experiment we \textit{post-select the events in which the photons exit different spatial modes}, retaining only the cross terms
\begin{equation}
|\psi\rangle_{\text{post}} \propto
t_H^\dagger r_V^\dagger
+
r_H^\dagger t_V^\dagger.
\end{equation}
Up to a normalization factor and a relative phase determined by the optical path lengths, this post-selected state spans the antisymmetric polarization subspace
\begin{equation}
|\psi^{-}\rangle =
\frac{1}{\sqrt{2}}
\left(
|H_t V_r\rangle
-
|V_t H_r\rangle
\right),
\end{equation}
which corresponds to the Bell singlet state used in the subsequent fusion protocol.}

\begin{figure}
\centering
\includegraphics[width=1.0\linewidth]{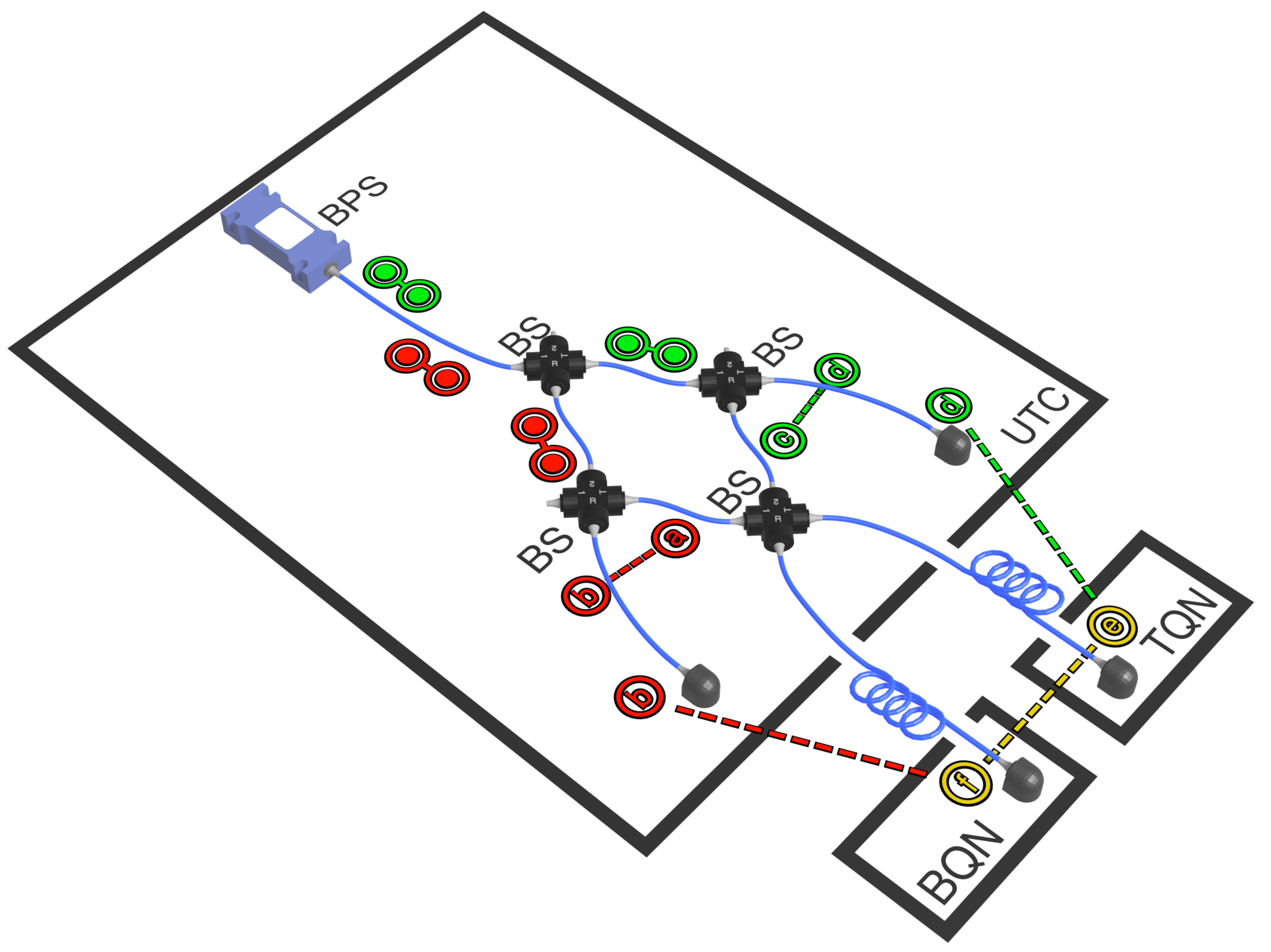}
\caption{Schematic illustrating our scheme for probabilistically splitting two spatially degenerate photon pairs and implementing a four-photon fusion operation across three nodes, labeled UTC, TQN, and BQN, using a series of fiber-coupled BSs. Two photons are retained locally at UTC for polarization-resolved heralding, while two photons are transmitted through deployed fiber to the two remote nodes of BQN and TQN as shown in Fig.~\ref{fig:Birdview}(b).
\label{fig:Layout}}
\end{figure}

\textcolor{black}{In the present experiment, two such probabilistically separated photon pairs are generated and routed through a network of beam splitters that implement a fusion operation, as shown in Fig.~\ref{fig:Layout}. Because each splitting event occurs with probability $1/2$, the overall probability of obtaining the desired four-photon configuration after the three beam splitters is $1/32$. The resulting four-photon state then serves as the input for the fusion-based interference process.}


As illustrated in Fig.~\ref{fig:Layout}, the SPDC source (labeled `BPS') generates two spatially degenerate photon pairs (denoted as the red and green pairs). After propagating through three successive BSs, the two pairs are probabilistically split into spatial modes \(a, b\) for the red pair and spatial modes \(c, d\) for the green pair, respectively. The resulting initial four-photon state, before \textit{arriving simultaneously} at the final BS, is given by the tensor product:
\begin{equation}
\begin{aligned}
|\psi\rangle_{\text{initial}}
= \frac{1}{2} \left( a_H^\dagger b_V^\dagger - a_V^\dagger b_H^\dagger \right)
\left( c_H^\dagger d_V^\dagger - c_V^\dagger d_H^\dagger \right) |0\rangle\\
= \frac{1}{2} ( |H_a V_b H_c V_d\rangle - |H_a V_b V_c H_d\rangle
\\
- |V_a H_b H_c V_d\rangle + |V_a H_b V_c H_d\rangle ).
\end{aligned}
\label{eqn:Ini_reframed}
\end{equation}
In the fusion stage, modes \(a\) and \(c\) are interfered on a 50/50 beam splitter to produce output modes \(e\) and \(f\), while modes \(b\) and \(d\) are retained for polarization-resolved conditioning.

A 50/50 BS transforms the creation operators of the input modes ($a, c$) into those of the output modes ($e, f$) according to:
\begin{equation}
a_p^\dagger \to \frac{1}{\sqrt{2}} \left( e_p^\dagger + f_p^\dagger \right), \quad
c_p^\dagger \to \frac{1}{\sqrt{2}} \left( e_p^\dagger - f_p^\dagger \right),
\label{eqn:Trans}
\end{equation}
where $p \in \{H,V\}$. Modes $b$ and $d$ are unaffected by the final BS.

Applying Eq.~(\ref{eqn:Trans}) to the four-photon components in Eq.~(\ref{eqn:Ini_reframed}) yields a superposition over output-number patterns in modes \(e\) and \(f\), including (i) coincidence terms with one photon in each output mode, and (ii) bunching terms with two photons in a single output mode. When the interfering photons are indistinguishable in all degrees of freedom, multi-photon interference suppresses particular coincidence pathways (Hong--Ou--Mandel--type suppression), producing a dip in appropriately chosen four-fold coincidence counts as the relative temporal delay is scanned. Explicitly, the resulting four-photon state can be written as follows:
\begin{equation}
\begin{aligned}
|\psi\rangle_{\text{final}} = \frac{1}{2\sqrt{2}} \Big[ \left( |2_H\rangle_e - |2_H\rangle_f \right) |V_b V_d\rangle  \\
- \frac{1}{\sqrt{2}}\left( |H V\rangle_e - |H\rangle_e |V\rangle_f + |V\rangle_e |H\rangle_f - |H V\rangle_f \right) |V_b H_d\rangle  \\
- \frac{1}{\sqrt{2}}\left( |V H\rangle_e - |V\rangle_e |H\rangle_f + |H\rangle_e |V\rangle_f - |V H\rangle_f \right) |H_b V_d\rangle \\
+ \left( |2_V\rangle_e - |2_V\rangle_f \right) |H_b H_d\rangle \Big].
\end{aligned}
\label{final_BS}
\end{equation}
Therefore, by controlling the polarizations of the two unaffected/locally retained photons in modes \( b \) and \( d \), and performing four-photon coincidence measurements, the final state described by Eq.~(\ref{final_BS}) can be projected onto either coincidence channels consistent with N00N-like correlation (mode-mode correlation) or a Bell-like correlation  (particle-particle correlation) shared between modes \( e \) and \( f \). Specifically, when photons in modes \( b \) and \( d \) are identically polarized (both \( H \) or both \( V \)), the resulting state in modes \( e \) and \( f \) is a N00N state of the form \( \ket{2_V}_e - \ket{2_V}_f \) or \( \ket{2_H}_e - \ket{2_H}_f \), respectively. In contrast, when photons in \( b \) and \( d \) are cross-polarized, the \textit{post-selected four-photon coincidence} projects modes \( e \) and \( f \) into the Bell singlet state \( \ket{H}_e \ket{V}_f - \ket{V}_e \ket{H}_f \). 

As illustrated in Fig.~\ref{fig:Layout}, the two unaffected photons in modes \( b \) and \( d \) are retained locally at UTC, while the photons in modes \( e \) and \( f \) are transmitted to two separate nodes on the network, labeled `BQN' and `TQN' in the diagram, which are also shown in the bird's-eye view in Fig.~\ref{fig:Birdview}(b). It therefore constitutes a \textit{four-photon interference shared among three spatially separated nodes} on a deployed fiber network infrastructure.

In this work, we use this polarization-controlled reconfigurable conditioning to switch between these two interference regimes in the deployed fiber network and quantify the resulting changes in multi-photon coincidence statistics. Throughout, we refer to these outcomes as \textit{Bell-like} and \textit{N00N-like} conditional channels, and we reserve the term \textit{entanglement} for future work that includes explicit multi-partite entanglement witnesses. 

\subsection{Polarization-Correlation Fringe Model for the Bell-like Channel}

For completeness, we summarize the expected polarization-correlation dependence for the Bell-like conditional channel in the idealized case where a singlet-like two-photon component is selected. An antisymmetric two-photon polarization state with a relative phase delay \( \Delta\phi \) between orthogonal polarizations can be written as
\begin{equation}
\ket{\psi^-} = \frac{1}{\sqrt{2}} \left( \ket{HV} - e^{i\Delta\phi} \ket{VH} \right),
\end{equation}
where \( \Delta\phi \), in our experiments, is introduced by a liquid crystal variable retarder (LCVR). By rotating the slow axis of the input polarization-maintaining (PM) fiber connected to the two polarizing beam splitters (PBSs) by \( 45^\circ \) relative to the horizontal axis of the lab frame (labeled as the red-colored input fiber in Fig.~\ref{fig:Source}(b)), the state is expressed in the diagonal/anti-diagonal \( \{ \ket{D}, \ket{A} \} \) basis as:
\begin{equation}
\ket{\psi^-} = \frac{1}{\sqrt{2}} \left( \ket{DA} - e^{i\Delta\phi} \ket{AD} \right).
\label{eqn:AD}
\end{equation}
With
\begin{equation}
\ket{D} = \frac{1}{\sqrt{2}} \left( \ket{H} + \ket{V} \right), \quad
\ket{A} = \frac{1}{\sqrt{2}} \left( \ket{H} - \ket{V} \right),
\end{equation}
one obtains the idealized coincidence probabilities
\begin{eqnarray}
\nonumber
P_{HH,VV} = \frac{1}{4} (1 - \cos \Delta\phi), \\
P_{HV,VH} = \frac{1}{4} (1 + \cos \Delta\phi).
\label{eqn:Fit}
\end{eqnarray}
In the present work, we use these expressions as a convenient reference model to compare against measured polarization-correlation fringes, while emphasizing that entanglement verification would require additional measurements (e.g., an entanglement witness or Bell test) beyond the scope of this field demonstration.

\section{Experiments and Results}

\subsection{Characterization of the Fiber-Coupled Bi-Photon Source}

\begin{figure}
\centering
\includegraphics[width=1.0\linewidth]{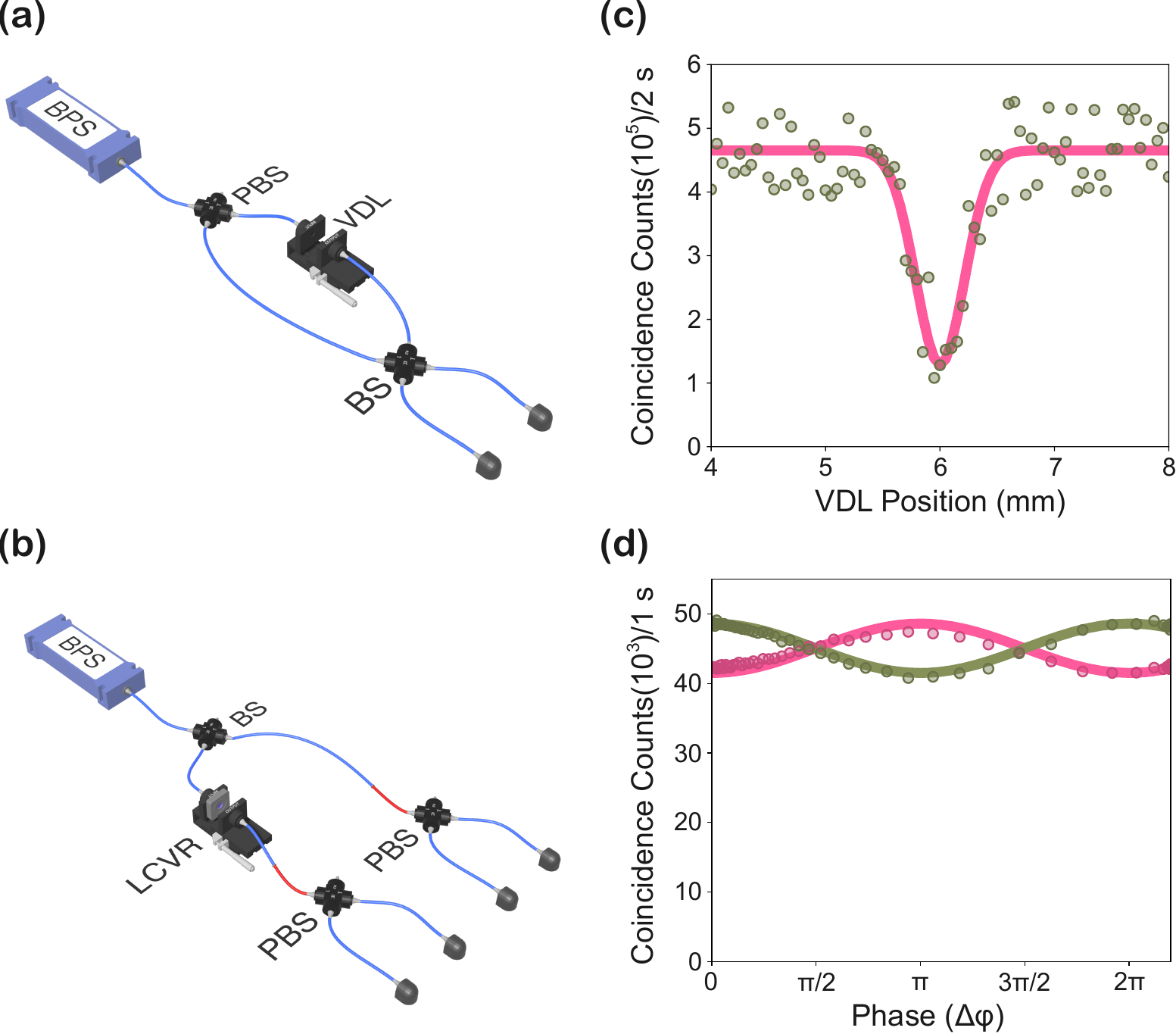}
\caption{Characterization of the fully fiber-coupled spatially-degenerate BPS through (a) HOM interference and (b) polarization-correlation fringe measurements with an LCVR, with the corresponding experimental results shown in (c) and (d), respectively. The red-colored PM fibers connected to the two PBSs in (b) indicates that their slow axes are rotated by $45^\circ$ relative to the horizontal axis of the lab frame at the input of the two PBSs. Photons are detected using SNSPDs, and two-photon coincidences are recorded with a coincidence window of 8~ns. The solid pink line in (c) is a Gaussian fit, while the solid green and pink lines in (d) are theoretical curves for the two-photon coincidence probabilities \( P_{HH} \) and \( P_{HV} \) respectively from Eq.~(\ref{eqn:Fit1}).
\label{fig:Source}}
\end{figure}

In this work, we employ an all-fiber-optic platform consisting of a continuous-wave (CW) fully fiber-coupled bi-photon source (manufactured by Qubitekk and labeled `BPS') and fiber-coupled linear optics, including BSs, PBSs, and a variable delay line (VDL) (all manufactured by OZ Optics). The BPS is based on type-II SPDC in periodically-poled KTP crystal and is temperature-tuned to produce photon pairs near 1570~nm with a bandwidth of approximately 3~nm. Because the source output is fully fiber-coupled and spatially degenerate, the photons are separated and routed using fiber BSs as shown in Fig.~\ref{fig:Layout}. We characterize the source using (i) Hong--Ou--Mandel (HOM) interference~\cite{hong1987measurement} and (ii) polarization-correlation fringe measurements based on the phase-dependent model in Eq.~(\ref{eqn:Fit}). The corresponding experimental schematics are shown in Figs.~\ref{fig:Source}(a) and~\ref{fig:Source}(b).

\textcolor{black}{Throughout this work, the slow axis of the PM fiber connected to the reflected output of the PBS is rotated so that both output photons are \( H \)-polarized before detection by superconducting nanowire single-photon detectors (SNSPDs), maximizing detection efficiency. As shown in Fig.~\ref{fig:Source}(a), temporal overlap at the BS is achieved by scanning the VDL. The characterization data in Figs.~\ref{fig:Source}(c) and~\ref{fig:Source}(d) were acquired using a two-photon coincidence window of 8~ns. The Gaussian fit in Fig.~\ref{fig:Source}(c) yields a HOM visibility, \( V_{\mathrm{HOM}} = (C_{\mathrm{max}} - C_{\mathrm{min}})/C_{\mathrm{max}} \)~\cite{ramesh2025hong}, of 68~\%. 
The accidental coincidence rate is estimated to be \(2\times10^{3}\,\mathrm{s^{-1}}\), which is sufficiently small to have a negligible effect on the measured visibility.
The curves in Fig.~\ref{fig:Source}(d) are theoretical predictions from Eq.~(\ref{eqn:Fit1}) after mapping LCVR voltage to \( \Delta\phi \) using the wavelength--voltage calibration data provided in the Thorlabs datasheet (LCC1613-C)~\cite{datasheet}.}

\textcolor{black}{The elevated background observed in the polarization-correlation fringes can be quantitatively understood by modeling the emitted photon pairs as a mixture of a Bell-singlet component and a phase-independent background. Writing the density matrix as \( \rho = p|\psi^{-}\rangle\langle\psi^{-}| + (1-p)\rho_{\mathrm{bg}} \), where \(p\) denotes the fraction of photon pairs occupying the coherent Bell-like polarization subspace, the coincidence probabilities after the two \(45^\circ\)-rotated PBS analyzers become 
\begin{eqnarray}
\nonumber
P_{HH,VV} = \frac{1}{4} (1 - p\cos \Delta\phi), \\
P_{HV,VH} = \frac{1}{4} (1 + p\cos \Delta\phi).
\label{eqn:Fit1}
\end{eqnarray}
The oscillatory component therefore has amplitude \(p/4\), while the constant background remains approximately \(1/4\), so the interference appears as a small fringe on top of a large coincidence pedestal. From the measured fringe contrast we estimate \(p\approx0.17\), implying that only about~\(17~\%\) of detected photon pairs contribute to the coherent interference process.}

\textcolor{black}{This observation is consistent with the independently measured HOM visibility \(V_{\mathrm{HOM}}\approx0.68\), which provides an estimate of the effective photon indistinguishability of the source 
For SPDC sources the HOM visibility can be approximated as \( V_{\mathrm{HOM}} \approx (1/K)\cdot({1+\mu)^{-1}}\)~\cite{Grice2001,Mosley2008,Christ2012},
where \(K\) is the Schmidt number describing the spectral purity of the biphoton wavefunction and \(\mu\) is the mean pair-generation probability within the coincidence window. Using the measured pair rate \(R_{\mathrm{pair}}\approx2.4\times10^{5}\,\mathrm{s^{-1}}\) and the 8~ns coincidence window used in the two-photon measurements, we estimate \(\mu \approx R_{\mathrm{pair}}\Delta t \approx (2.4\times10^{5})\cdot(8\times10^{-9}) \approx 1.9\times10^{-3}\). Since \(\mu \ll 1\), the factor \(1/(1+\mu)\approx1\), indicating that multi-pair emission within the coincidence window plays only a minor role in limiting the observed HOM visibility. }

\textcolor{black}{The measured value \(V_{\mathrm{HOM}}\approx0.68\) therefore primarily reflects residual spectral impurity of the biphoton state, corresponding to an effective Schmidt number \(K\approx1/0.68\approx1.47\). The fact that \(p \ll V_{\mathrm{HOM}}\) further indicates that only a subset of the partially indistinguishable photon pairs remains in the Bell-like polarization subspace after propagation through the BSs and polarization-analysis optics. This effect likely arises from the \(45^\circ\)-rotated PM-fiber slow-axis misalignment and polarization mismatch across fiber mating sleeves. The remaining photon pairs contribute a phase-independent coincidence background, which produces the elevated offset observed in the experimental data.}


\textcolor{black}{We emphasize that the source is a commercial, closed, fully fiber-coupled unit with no available alignment degrees of freedom for optical performance optimization. 
We therefore focus on \textit{robust, qualitative signatures of multi-photon interference}, and \textit{polarization-controlled reconfigurable conditional multi-state projection} in the deployed network.}

\subsection{Multi-Photon Coincidence Measurements Across the Network}

\begin{figure}
\centering
\includegraphics[width=1.0\linewidth]{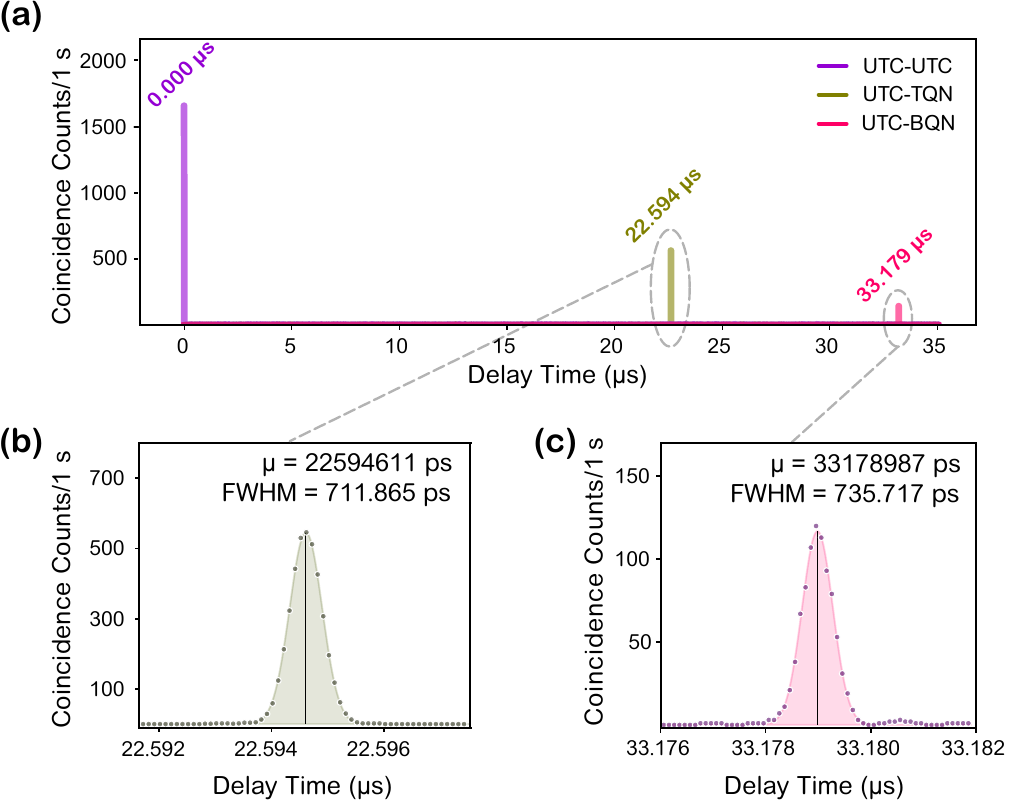}
\caption{
Characterization of the relative photon arrival times across the network. (a) Relative arrival times of the four photons, with one photon sent to TQN and another one to BQN, while the rest two are retained at UTC. (b)\&(c) Zoomed-in views of the temporal correlation measurements between UTC–TQN and UTC–BQN, respectively. The relative delay between the two photons retained locally at UTC, represented by the purple line, is too small to be resolved in (a). 
\label{fig:Sync}}
\end{figure}

Since photons are sent to distant nodes and loop back, precise time tagging is essential for implementing multi-photon coincidence measurements. We gate detection events in software using measured timing correlations, which suppresses background noise (dominated by spontaneous Raman scattering when single-photon-level quantum signals co-propagate with wavelength-multiplexed classical timing pulses) and reduces accidental coincidences. To determine relative photon arrival times, we build coincidence histograms 
between UTC and TQN and between UTC and BQN using an  Hanbury–Brown–Twiss (HBT)–type configuration in which one photon is retained at UTC and the other is sent to a remote node. The results are shown in Fig.~\ref{fig:Sync}(a), with zoomed-in timing correlations in Figs.~\ref{fig:Sync}(b) and~\ref{fig:Sync}(c). We determine the relative delays to be 22.594~\(\mu\)s (UTC--TQN) and 33.179~\(\mu\)s (UTC--BQN). Given a timing-correlation FWHM of approximately 750~ps, we select a coincidence window of 2~ns to capture correlated events while accommodating possible arrival-time drift due to environmental disturbances during long acquisitions.

The EPB--IonQ network introduces attenuation due to all-optical switching used for \textit{reconfigurable routing}~\cite{earl2022architecture}: approximately 5~dB round-trip loss between UTC and TQN and approximately 10~dB round-trip loss between UTC and BQN. While such attenuation can be tolerated in some loss-robust networking protocols~\cite{bersin2024development,crum2024clock}, it substantially reduces four-fold coincidence rates in our multi-photon interference experiments. In the following sections we report the resulting acquisition times and discuss practical implications for scaling to higher-rate field demonstrations.

\subsection{Fusion Operation Across the Network}

\begin{figure}
\centering
\includegraphics[width=1.0\linewidth]{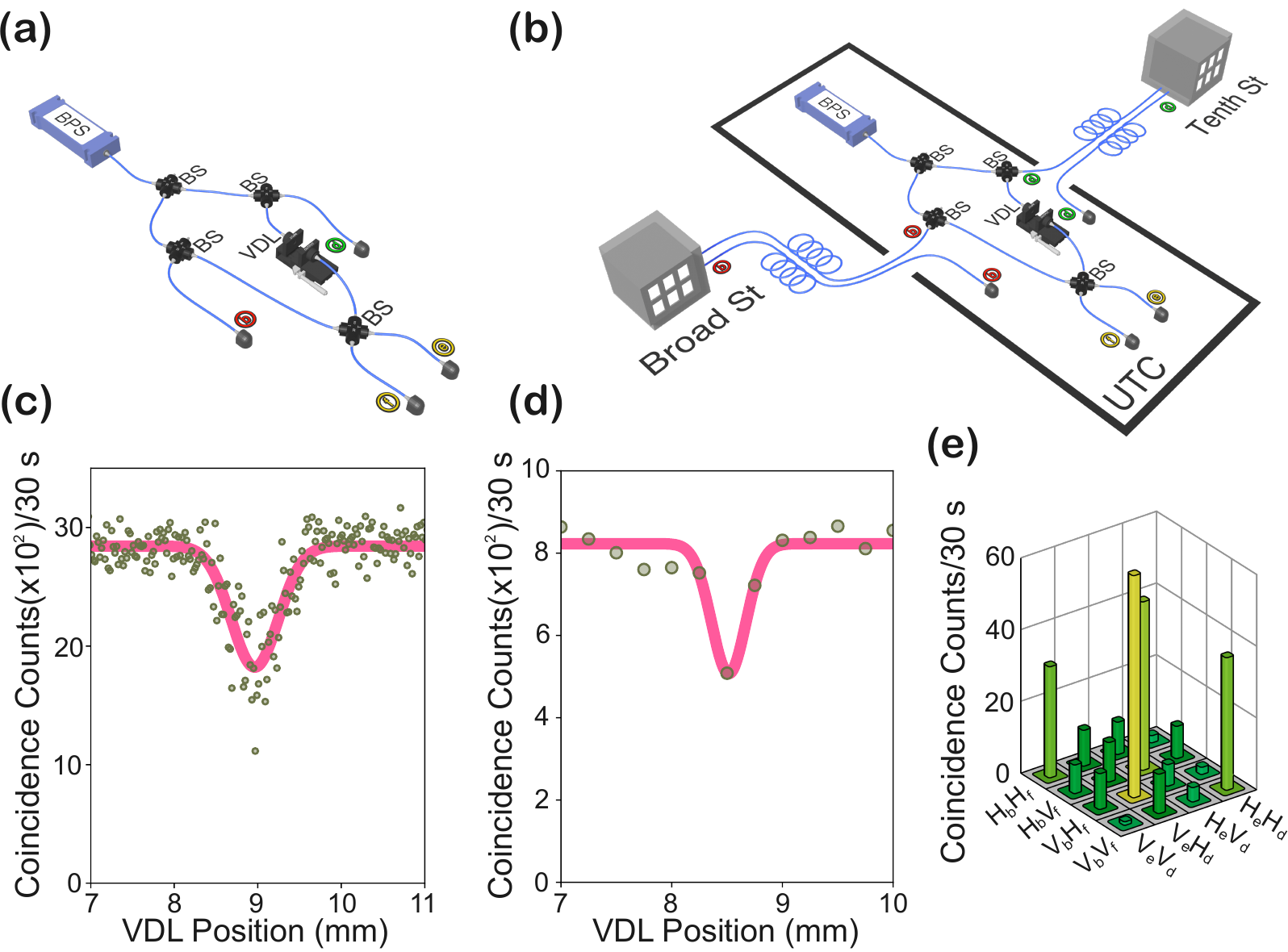}
\caption{(a) Experimental layout for the four-photon fusion stage locally at UTC. (b) Experimental layout for the four-photon fusion stage across the network. \textcolor{black}{In both (a)\&(b), the propagation and detection paths of the four photons are labeled \(b\), \(d\), \(e\), and \(f\), consistent with the experimental schematic shown in Fig.~\ref{fig:Layout}. In the network setting (b), photons \(e\) and \(f\) are retained at UTC, while photons \(b\) and \(d\) are routed through deployed fiber loops to TQN and BQN, respectively, before returning for detection.} The solid circles in (c)\&(d) are the experimental results corresponding to (a)\&(b), respectively. A dip in the four-fold coincidence counts indicates multi-photon interference at the fusion BS. (e) Four-fold polarization-resolved coincidence counts over the 16 
measurement outcomes, recorded at the dip setting of (d). The solid pink lines in (c)\&(d) are Gaussian fits. 
\label{fig:Fusion}}
\end{figure}

With the source and arrival-time synchronization characterized, we demonstrate the four-photon fusion stage described in Section~III.A across the network. To highlight network reconfigurability, in this experiment we retain the photons in modes \( e \) and \( f \) at UTC while routing modes \( b \) and \( d \) to BQN and TQN, respectively (Fig.~\ref{fig:Fusion}(b)). We first perform the fusion operation locally by retaining all four photons at UTC (Fig.~\ref{fig:Fusion}(a)). The local and network fusion results are shown in Figs.~\ref{fig:Fusion}(c) and~\ref{fig:Fusion}(d), respectively. As temporal overlap is achieved by adjusting the VDL, multi-photon interference suppresses distinguishable-path coincidence contributions, producing a dip in the measured four-fold coincidence rate. We interpret these dips as evidence of successful temporal overlap and indistinguishability sufficient to observe fusion-stage interference in both local and deployed-fiber configurations.

To further characterize the polarization-dependent structure of the four-fold coincidence events at the network-fusion setting, we fix the VDL at the dip position in Fig.~\ref{fig:Fusion}(d) and record four-fold coincidences across all \(2^4=16\) polarization outcomes (Fig.~\ref{fig:Fusion}(e)). The corresponding four-photon coincidence counts are shown in Fig.~\ref{fig:Fusion}(e), with four pronounced peaks for \( H_b H_f V_e V_d \), \( V_b V_f H_e H_d \), \( H_b V_f H_e V_d \), and \( V_b H_f V_e H_d \), confirming the presence of the four-photon (\( e, f, b, d \)) components in Eq.~(\ref{final_BS}). These measurements confirm the presence of the expected four-photon coincidence components in the post-selected detection subspace and serve as a consistency check for subsequent polarization-controlled reconfigurable conditional measurements. 

\subsection{Bell-like Conditional Channel Across the Network}

\begin{figure}
\centering
\includegraphics[width=1.0\linewidth]{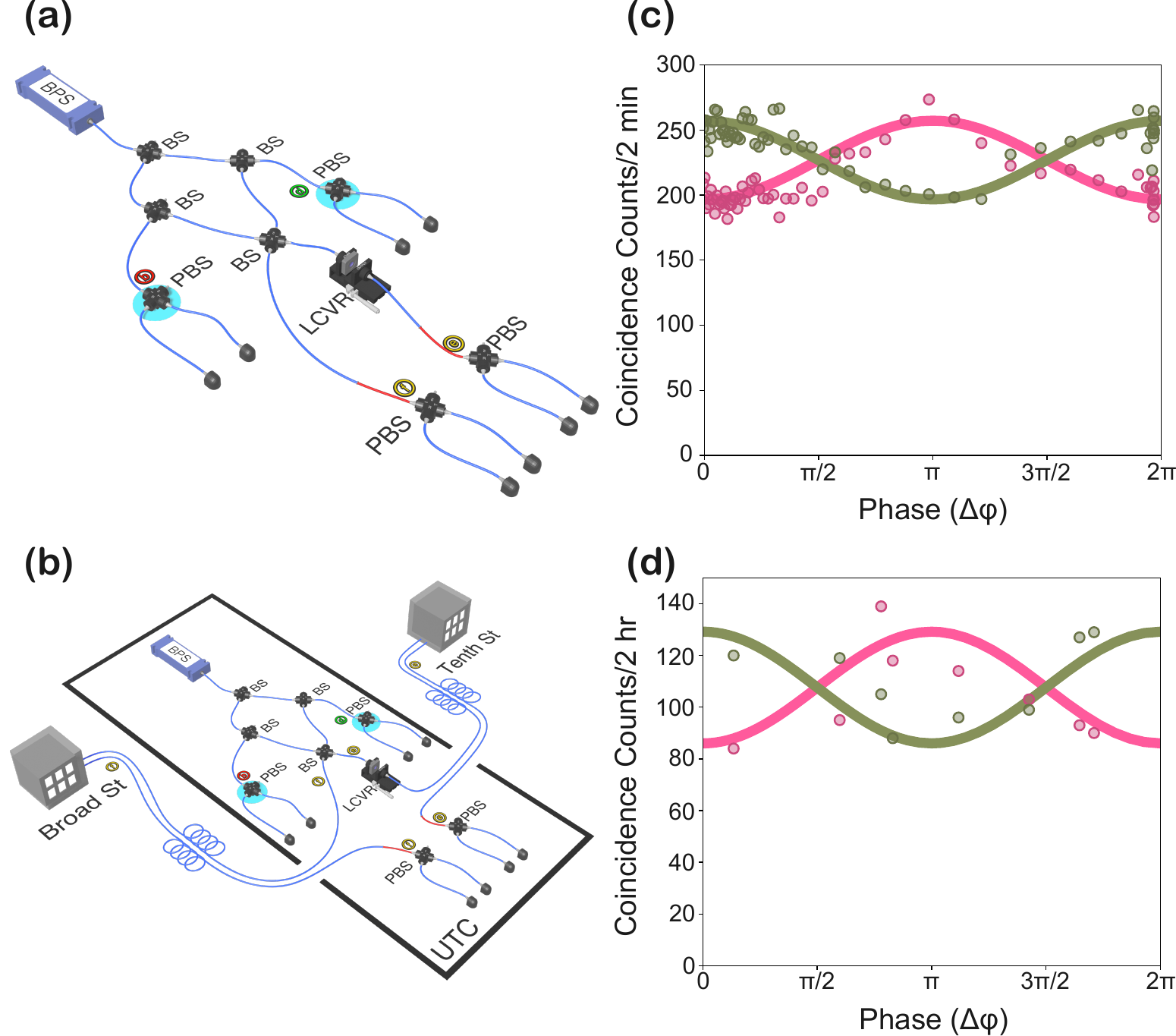}
\caption{(a) Experimental layout for polarization-controlled generation of the Bell-like conditional channel locally at UTC. (b) Experimental layout for the operation across the network. \textcolor{black}{In both (a)\&(b), the propagation and detection paths of the four photons are labeled \(b\), \(d\), \(e\), and \(f\), consistent with the experimental schematic shown in Fig.~\ref{fig:Layout}. In the network setting (b), photons \(b\) and \(d\) are retained at UTC for polarization control, while photons \(e\) and \(f\) are routed  through deployed fiber loops to TQN and BQN, respectively, before returning for detection.} Data shown in (c)\&(d) correspond to (a)\&(b), respectively. The solid green and pink circles in (c)\&(d) represent the measured four-photon coincidence counts $C_{H_e H_f}\big|_{V_b H_d}$ and $C_{H_e V_f}\big|_{V_b H_d}$, respectively. The thick solid green and pink curves in panels (c)\&(d) denote the theoretical curves for the coincidence probabilities $P_{HH}$ and $P_{HV}$ from Eq.~(\ref{eqn:Fit}), respectively, while the thinner green and pink lines are provided as visual guides for the experimental data.
\label{fig:Bell}}
\end{figure}

We next demonstrate polarization-controlled reconfigurable conditional projection corresponding to a Bell-like coincidence channel. We retain the two conditioning photons in modes \( b \) and \( d \) locally at UTC and route modes \( e \) and \( f \) through deployed fiber loops to TQN and BQN, respectively, before returning for detection. By projecting the locally retained photons \( b \) and \( d \) into cross-polarized outcomes (e.g., \( V_b H_d \) or \( H_b V_d \)) using the PBSs highlighted in cyan in Figs.~\ref{fig:Bell}(a) and~\ref{fig:Bell}(b), and by recording four-fold coincidences, we select a conditional subset of remote events whose polarization-correlation dependence follows the theoretical model in Eq.~(\ref{eqn:Fit}). 
The local and deployed-network configurations are shown in Figs.~\ref{fig:Bell}(a) and~\ref{fig:Bell}(b), with the corresponding results presented in Figs.~\ref{fig:Bell}(c) and~\ref{fig:Bell}(d).

For the data shown, the conditioning two photons in modes \( b \) and \( d \) are detected first, defining the conditional subset for the subsequent detections associated with the two photons traversing the deployed fiber loops. Although the network data are noisier due to environmental disturbances and loss, the overall polarization-dependent four-fold coincidence trends remain visible. We emphasize that these measurements do not exceed standard inseparability thresholds and are therefore presented as polarization-controlled \textit{Bell-like} conditional correlation signatures on a deployed network.


\subsection{N00N-like Conditional Channel Across the Network}

\begin{figure}
\centering
\includegraphics[width=1.0\linewidth]{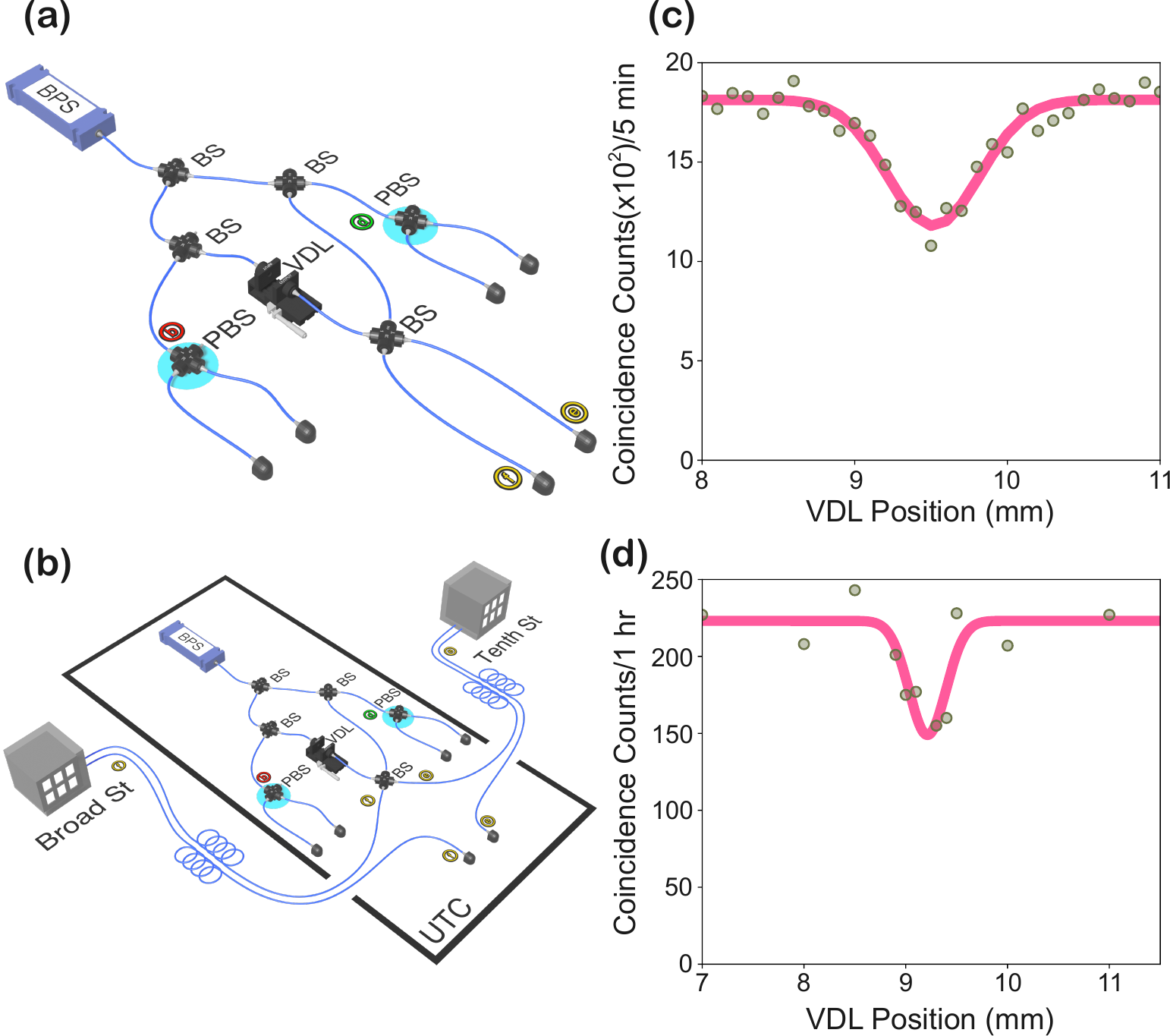}
\caption{(a) Experimental layout for polarization-controlled operation of the N00N-like conditional channel locally at UTC. (b) Experimental layout for the operation across the network. \textcolor{black}{In both (a)\&(b), the propagation and detection paths of the four photons are labeled \(b\), \(d\), \(e\), and \(f\), consistent with the experimental schematic shown in Fig.~\ref{fig:Layout}. In the network setting (b), photons \(b\) and \(d\) are retained at UTC for polarization control, while photons \(e\) and \(f\) are routed  through deployed fiber loops to TQN and BQN, respectively, before returning for detection.} Data shown in (c)\&(d) corresponds to (a)\&(b), respectively. The solid green circles in (c)\&(d) represent the measured four-photon coincidence counts with photons \( b \) and \( d \) projected into the same-polarization modes of \( H_b H_d \). The thick solid pink lines in (c)\&(d) are Gaussian fits, while the thinner green lines are provided as
visual guides for the experimental data.
\label{fig:NooN}}
\end{figure}

Finally, we demonstrate switching to a N00N-like conditional channel by projecting the locally retained photons in modes \( b \) and \( d \) onto same-polarization outcomes (e.g., \( H_b H_d \) or \( V_b V_d \)) using the PBSs highlighted in cyan in Figs.~\ref{fig:NooN}(a) and~\ref{fig:NooN}(b). Under this conditioning, Eq.~(\ref{final_BS}) collapses into a N00N state of the form \( \ket{2_V}_e - \ket{2_V}_f \) or \( \ket{2_H}_e - \ket{2_H}_f \), thereby projecting TQN and BQN into a \textit{mode-mode} (or \textit{path-path}) correlation, as opposed to the \textit{particle-particle} correltion of a Bell pair. The experimental layouts for the N00N-like conditional channel generation locally and distributed across the network to TQN and BQN are shown in Figs.~\ref{fig:NooN}(a) and~\ref{fig:NooN}(b), respectively. By measuring the four-photon coincidences with photons \( b \) and \( d \) projected into the same-polarization modes (for the data shown, they are in \( H_b H_d \) modes), the post-selected four-fold coincidence subset emphasizes same-mode (bunching) contributions in the \(e/f\) outputs, consistent with the expected behavior of a N00N-like mode-superposition channel in the theoretical model. As in the fusion measurements, scanning the VDL reveals a dip in the four-fold coincidence counts when temporal overlap at the fusion BS is achieved as shown in Figs.~\ref{fig:NooN}(c) and~\ref{fig:NooN}(d).

The observed visibility is suboptimal, with elevated dip values, consistent with the source and interferometric limitations discussed in Section~IV.A. Also note that the most direct way to quantitatively characterize a N00N state is by measuring its Heisenberg-limited phase sensitivity~\cite{wang2025exact,kim2025distributed}. While our current network configuration and count rates do not allow such a measurement within reasonable acquisition times, this will be the focus of a dedicated future study.

\textcolor{black}{Also note that the purpose of the measurements in Figs.~\ref{fig:Fusion}(d),~\ref{fig:Bell}(d),~\ref{fig:NooN}(d)  is to demonstrate the presence of the four-photon interference signature under field-deployed network conditions rather than to extract a physical parameter with high statistical precision. The dominant fluctuations arise from slow systematic variations of the deployed fiber network (e.g., polarization drift, fiber stress, temperature variation, and switch routing variation) \textit{rather than from Poisson counting statistics}. \textit{Consequently, statistical error bars derived solely from photon-counting noise would not accurately represent the true experimental uncertainty}. Furthermore, the four-photon coincidence rate in the deployed network is extremely low (on the order of $10^{-2}$--$10^{-1}\,\mathrm{Hz}$), as shown in  Figs.~\ref{fig:Bell}(d) and~\ref{fig:NooN}(d), acquiring statistically independent repetitions for each data point would require measurement times of several days per interference curve. Instead, the interference visibility is extracted from a global fit to the measured interference pattern, which effectively averages over statistical fluctuations across the dataset.
}

\section{Conclusion and Discussion}

\textcolor{black}{In summary, we have demonstrated polarization-controlled reconfigurable four-photon interference among three nodes and conditional two-photon projection over a metropolitan-scale deployed fiber network using an all-fiber linear-optical platform. By synchronizing four-photon arrival times across the three distant nodes on the deployed network and performing polarization-resolved conditioning on locally retained photons, we observe distinct four-fold coincidence interference signatures corresponding to Bell-like and N00N-like conditional channels. These results extend multi-photon interference experiments beyond the laboratory and into a real-world, multi-node telecommunication infrastructure characterized by substantial loss, polarization drift, and environmental perturbations.}

\textcolor{black}{From a systems perspective, this work serves two primary roles. First, it provides a field validation of multi-photon interference across multiple distant nodes as a viable network-level resource. Many advanced quantum-network protocols explicitly rely on the ability to generate and interfere multiple photons with sufficient indistinguishability~\cite{park2021four,huang2011experimental,pan2001experimental}, yet experimental evidence for such processes in deployed fiber has remained limited. Our results show that four-photon interference signatures can be observed across three distant nodes on a metropolitan network, while also quantifying the practical limitations imposed by source infidelity, fiber loss, and timing uncertainty.}

\textcolor{black}{Secondly, the polarization-controlled switching between Bell-like and N00N-like conditional coincidence channels demonstrates a \textit{programmable network control primitive} based entirely on linear optics and local measurements. This capability enables systematic benchmarking of deployed fiber links under different correlation structures and provides a pathway toward adaptive quantum network operation, where conditional multi-photon processes can be selected dynamically in response to network conditions. Notably, \textit{these capabilities remain meaningful even in the absence of entanglement verification}, making them directly relevant to near-term quantum network development.}

The current performance limitations are dominated by (i) source quality (limited polarization-correlation contrast and reduced indistinguishability) and (ii) network loss introduced by reconfigurable all-optical switching~\cite{earl2022architecture}. Both limitations directly impact multi-photon interference visibility and four-fold coincidence rates. Near-term improvements include replacing the present spatially degenerate commercial BPS with a higher-fidelity source (commercial or custom Sagnac-based~\cite{bell2014experimental}) and reducing photon-routing loss through planned all-optical switching upgrades. With improved source fidelity and network transmission, future work will implement explicit entanglement witnesses for the conditional channels demonstrated here.

In this work, all photon-arrival times are ultimately detected at UTC after traversing the deployed fiber loops. Extending to nonlocal multi-photon coincidence detection at physically separated nodes with independently running clocks will require tighter timing synchronization (e.g., via the White Rabbit protocol)~\cite{moreira2009white,lipinski2011white}, and a network configuration supporting distributed time-tag correlation without centralized detection. Finally, integrating automatic polarization control protocols (e.g., as in Ref.~\cite{chapman2024continuous}) will significantly improve long-term stability of multi-photon interference in deployed fiber environments. These developments form the foundation for our next stage of experiments.

\section*{acknowledgments}
JEH, AFE, GSA, and TL acknowledge support from the U.S. National Science Foundation (NSF) through the ExpandQISE program under Award No. 2426699. JEH, AFE, and TL also acknowledge support from the NSF CCSS program under Award No. 2503630. KR and TL acknowledge support from the U.S. National Institute of Standards and Technology (NIST) through the CIPP program under Award No. 60NANB24D218. MMH and GS acknowledge support from the U.S. Department of Energy, Office of Science, Office of Advanced Scientific Computing Research, through the Quantum Internet to Accelerate Scientific Discovery Program under Field Work Proposal No. 3ERKJ381, and from the NSF under Award No. DGE-2152168.





\bibliography{MyReferences}
\end{document}